\documentstyle[12pt]{article}

\begin{document}

\renewcommand{\footnote}[1]{%
\def\thefootnote{\arabic{footnote})}
\footnotemark%
\footnotetext{#1}}
\newfont{\blackboard}{cmr10}
\newcommand{\Z}{\mbox{\blackboard\symbol{"5A}}}
\def\Res{\mathop{\rm Res}}
\newcommand{\io}{[\hspace{-1pt}[}
\newcommand{\ic}{]\hspace{-1pt}]}
\newcommand{\fo}{\{\!\mid\!}
\newcommand{\fc}{\!\mid\!\}}
\renewcommand{\sb}{\mbox{\boldmath $\sigma$}}
\newcommand{\sgn}{{\rm sgn}}
\newcommand{\BS}{{\rm BS}}
\renewcommand{\Re}{{\rm Re\,}}
\newcommand{\REG}{{\rm REG}}
\newcommand{\IRREG}{{\rm IRREG}}
\newcommand{\mod}{{\rm mod}}
\newcommand{\ren}{{\rm ren}}
\newcommand{\tr}{{\rm tr}}
\renewcommand{\arctan}{{\rm arctan}}
\newcommand{\tg}{{\rm tg}}
\renewcommand{\cosh}{{\rm cosh}}
\newcommand{\qm}{({q\over|m|}}
\newcommand{\r}{\rangle}
\newcommand{\Lp}{L_{(+)}}
\newcommand{\Lm}{L_{(-)}}

\def\bea{\begin{eqnarray}}
\def\eea{\end{eqnarray}}
\def\ba{\begin{array}}
\def\ea{\end{array}}
\def\V{{\bf V}}
\def\J{{\bf J}}
\def\j{{\bf j}}
\def\v{{\bf v}}
\def\vac{{\rm vac}}
\def\x{{\bf x}}
\def\F{\Phi^{(0)}}
\def\M{{\cal M}}
\def\N{{\cal N}}
\def\S{{\cal S}}
\def\L{{\cal L}}
\def\A{{\cal A}}
\def\C{{\cal C}}
\def\il{{\int\limits}}
\def\i{{\int}}
\def\P{P^{-{1\over 2}-z}_{-{1\over 2}-z}}
\def\Pn{P^{-{1\over 2}+N}_{-{1\over 2}+N}}
\newcommand{\bpar}{\mbox{\boldmath $\partial$}}
\newcommand{\ab}{\mbox{\boldmath $\alpha$}}
\newcommand{\gb}{\mbox{\boldmath $\gamma$}}
\def\Sup{\mathop{\rm Sup}}
\def\si{\mathop{\displaystyle\sum\mkern-22mu\int\,}}

\newcommand{\ds}{\displaystyle}
\newcommand{\sss}{\scriptscriptstyle}

\setcounter{equation}{0}
\begin{center}

{\bf  Effects of singular external fields and boundary condition 
on the vacuum of massless fermions in QFT}

\bigskip

Yu. A. Sitenko

\medskip
 {\it Bogolyubov Institute for Theoretical Physics, National Academy of Sciences,}\\
 \medskip
 {\it Kyiv, 03143, Ukraine}

\end{center}

\begin{abstract}
Effects of the configuration of an external static magnetic field in
the form of a singular vortex on the vacuum of a quantized massless
spinor field are studied. The most general boundary conditions at
the punctured singular point which make the twodimensional Dirac
Hamiltonian to be self-adjoint are employed.
\end{abstract}

\section{Introduction}

A study of effects of singular external fields (zero-range potentials)
in quantum mechanics has a long history and has been comprehensively
conducted (see \cite{Alb} and references therein). Contrary to this,
effects of singular external fields in quantum field theory are at the
initial stage of consideration, and much has to be elucidated. Singular
background can act on the vacuum of a second-quantized spinor field in a
rather unusual manner: a leak of quantum numbers from the
singularity point occurs. This is due, apparently, to the fact that a
solution to the Dirac equation, unlike that to the Schrodinger one,
does not obey a condition of regularity at the singularity point. It is
necessary then to specify a boundary condition at this point, and the
least restrictive, but physically acceptable, condition is such that
guarantees self-adjointness of the Dirac Hamiltonian. Thus, effects of
polarization of the vacuum by a singular background appear to
depend on the choice of the boundary condition at the singularity
point, and the set of permissible boundary conditions is labelled, most
generally, by a self-adjoint extension parameter.

In this talk we shall consider some effects of 
polarization of the massless fermionic vacuum in the background of 
a pointlike magnetic vortex in 2+1-dimensional space-time 
(see also \cite{Sit01, Sit02}).

\section{Boundary Condition at the Location of the Vortex}

The operator of the second-quantized spinor field is presented in the
form
\begin{equation}
\Psi(\x,t)=\si_{E_\lambda>0}\, e^{-iE_\lambda
t}<\x|\lambda>a_\lambda+\si_{E_\lambda<0}\, e^{-iE_\lambda
t}<\x|\lambda>b_\lambda^+, 
\end{equation}
where $a_\lambda^+$ and $a_\lambda$ ($b_\lambda^+$ and $b_\lambda$) are
the spinor particle (antiparticle) creation and annihilation operators
satisfying the anticommutation relations,
and $<\x|\lambda>$ is the solution to the stationary Dirac equation
\begin{equation}
H<\x|\lambda>=E_\lambda<\x|\lambda>, 
\end{equation}
$H$ is the Dirac Hamiltonian, $\lambda$ is the set of parameters
(quantum numbers) specifying a state, $E_\lambda$ is the energy of
a state; symbol $\si$ means the summation over discrete and the
integration (with a certain measure) over continuous values of
$\lambda$. Ground state $|\vac>$ is defined conventionally as
\begin{equation}
a_\lambda|\vac>=b_\lambda|\vac>=0. 
\end{equation}
In the case of quantization of a massless spinor field in the
background of a static vector field $\V(\x)$, the Dirac Hamiltonian
takes form
\begin{equation}
H=-i\ab[\bpar-i\V(\x)], 
\end{equation}
where $\ab=\gamma^0\gb$ and $\beta=\gamma^0$ 
($\gb$ and $\gamma^0$ are the Dirac $\gamma$ matrices). In 
2+1-dimensional space-time $(\x,t)=(x^1,x^2,t)$, the Clifford algebra
has two inequivalent irreducible representations which can be differed
in the following way:
\begin{equation}
i\gamma^0\gamma^1\gamma^2=s, \qquad s=\pm1.
\end{equation}
Choosing the $\gamma^0$ matrix in the diagonal form,
one gets
\begin{equation}
\gamma^0=\sigma_3,
\quad \gamma^1=
e^{{i\over2}\sigma_3\chi_s}i\sigma_1e^{-{i\over2}\sigma_3\chi_s},
\quad \gamma^2=
e^{{i\over2}\sigma_3\chi_s}is\sigma_2e^{-{i\over2}\sigma_3\chi_s},
\end{equation}
where $\sigma_1,\sigma_2$ and $\sigma_3$ are the Pauli matrices, 
and $\chi_1$ and
$\chi_{-1}$ are the parameters that are varied in interval
$0\leq\chi_s<2\pi$ to go over to the equivalent representations.

The configuration of external field $\V(\x)=(V_1(\x),V_2(\x))$ is
chosen as
\begin{equation}
V_1(\x)=-\Phi^{(0)}{x^2\over(x^1)^2+(x^2)^2}, \quad V_2(\x)=\F
{x^1\over(x^1)^2+(x^2)^2}, 
\end{equation}
which corresponds to the magnetic field strength in the form of a singular
vortex
\begin{equation}
\bpar\times\V(\x)=2\pi\F\delta(\x),
\end{equation}
where $\F$ is the total flux (in $2\pi$ units) of the vortex -- i.e.
of the thread that pierces the plane $(x^1,x^2)$ at the origin. 

A solution to the Dirac equation (2) with Hamiltonian (4) in
background (7) can be presented as
\begin{equation}
<\x|E,n>=\left(\ba{l}
f_n(r,E)e^{in\varphi}\\[0.2cm]
g_n(r,E)e^{i(n+s)\varphi}\\ \ea \right), \quad n\in\Z,
\end{equation}
where $\Z$ is the set of integer numbers, $r=\sqrt{(x^1)^2+(x^2)^2}$ and 
$\varphi=\arctan(x^2/x^1)$ are the polar coordinates, and the radial functions
$f_n$ and $g_n$ satisfy the system of equations
\begin{equation}
e^{-i\chi_s}[-\partial_r+s(n-\F)r^{-1}]f_n(r,E)=E g_n(r,E),$$
$$e^{i\chi_s}[\partial_r+s(n-\F+s)r^{-1}]g_n(r,E)=E f_n(r,E).
\end{equation}
When vortex flux $\F$ is integer, the requirement of square
integrability for wave function (9) provides its regularity
everywhere on plane $(x^1,x^2)$, rendering partial Dirac
Hamiltonians for every value of $n$ to be essentially self-adjoint.
When $\F$ is fractional, the same is valid only for $n\neq
n_0$, where
\begin{equation}
n_0=\io\F\ic+{1\over2}-{1\over2}s, 
\end{equation}
$\io u\ic$ is the integer part of quantity $u$ (i.e., the greatest
integer that is less than or equal to $u$). For $n=n_0$, each of the
two linearly independent solutions to system (10) meets the
requirement of square integrability. Any particular solution in this
case is characterized by at least one (at most both) of the radial
functions being divergent as $r^{-p}$ ($p<1$) for $r\rightarrow 0$. 
Therefore, contrary to the case of $n\neq n_0$, the partial Dirac 
Hamiltonian in the case of $n=n_0$ is not essentially self-adjoint. The 
Weyl - von Neumann theory of self-adjoint operators (see, e.g., 
Ref.\cite{Akhie}) has to be employed in order to consider the
possibility of a self-adjoint extension in the case of $n=n_0$. It can
be shown that the self-adjoint extension exists indeed and is
parametrized by one continuous real variable denoted in the following
by $\Theta$. Thus, the partial Dirac Hamiltonian in the case of $n=n_0$
is defined on the domain of functions obeying the condition
\begin{equation}
\cos\bigl(s{\Theta\over2}+{\pi\over4}\bigr)\lim_{r\rightarrow 0}(\mu
r)^Ff_{n_0}=-e^{i\chi_s}\sin\bigl(s{\Theta\over2}+{\pi\over4}\bigr)
\lim_{r\rightarrow 0}(\mu r)^{1-F}g_{n_0}, 
\end{equation}
where $\mu>0$ is the parameter of the dimension of inverse length and
\begin{equation}
F=s\fo\F\fc+{1\over2}-{1\over2}s,
\end{equation}
$\fo u\fc=u-\io u\ic$ is the fractional part of quantity $u$,
$0\leq\fo u\fc<1$; note here that Eq.(12) implies that $0<F<1$,
since, for $F={1\over2}-{1\over2}s$, both $f_{n_0}$ and
$g_{n_0}$ obey the condition of regularity at $r=0$. Note also that Eq.(12) 
is periodic in $\Theta$ with the period of $2\pi$; therefore,
without a loss of generality, all permissible values of $\Theta$ will be
restricted in the following to the range $-\pi\leq\Theta\leq\pi$.

All solutions to the Dirac equation in the background of a singular
magnetic vortex correspond to the continuous spectrum and, therefore,
obey the orthonormality condition
\begin{equation}
\int d^2x<E,n|\x><\x|E',n'>={\delta(E-E')\over\sqrt{|EE'|}}\delta_{nn'}.
\end{equation}
In the case of $0<F<1$ one gets the following expressions
corresponding to the regular solutions with $sn>sn_0$:
\begin{equation}
\left(\ba{c} f_n\\ g_n \\ \ea \right) ={1\over2\sqrt{\pi}}
\left(\ba{c}
J_{l-F}(kr)e^{i\chi_s}\\[0.2cm]
\sgn(E)J_{l+1-F}(kr)\\ \ea \right), \qquad l=s(n-n_0), 
\end{equation}
the regular solutions with $sn<sn_0$:
\begin{equation}
\left(\ba{c} f_n\\ g_n \\ \ea \right)
={1\over2\sqrt{\pi}}
\left(\ba{c}
J_{l'+F}(kr)e^{i\chi_s}\\[0.2cm]
-\sgn(E)J_{l'-1+F}(kr) \ea \right), \qquad l'=s(n_0-n), 
\end{equation}
and the irregular solution:
\begin{equation}
\left(\ba{c} f_{n_0}\\ g_{n_0} \\ \ea \right)
={1\over 2\sqrt{\pi[1+\sin(2\nu_E)\cos(F\pi)]}}\times$$
$$\times \left(\ba{c}
[\sin(\nu_E)J_{-F}(kr)+\cos(\nu_E)J_F(kr)]e^{i\chi_s}\\[0.2cm]
\sgn(E)[\sin(\nu_E)J_{1-F}(kr)-\cos(\nu_E)J_{-1+F}(kr)]\\ \ea \right);
\end{equation}
here $k=|E|$, $J_{\rho}(u)$ is the Bessel function of order
$\rho$ and
$$
\sgn(u)=\left\{\ba{cc}
1,& u>0\\
-1,& u<0\\ \ea \right\}.
$$
Substituting the asymptotic form of Eq.(17) at $r\rightarrow 0$
into Eq.(12), one arrives at the relation between the parameters
$\nu_E$ and $\Theta$:
\begin{equation}
\tan(\nu_E)=\sgn(E)\bigl({k\over2\mu}\bigr)^{2F-1}\,
{\Gamma(1-F)\over\Gamma(F)}\tan\bigl(s{\Theta\over2}+{\pi\over4}\bigr),
\end{equation}
where $\Gamma(u)$ is the Euler gamma function.

\section{Fermion Number}

In the second-quantized theory the
operator of the fermion number is given by the expression
\begin{equation}
\hat{\N}=\int d^2x{1\over2}[\Psi^+(\x,t),\Psi(\x,t)]_-=\si
[a_\lambda^+a_\lambda-b_\lambda^+b_\lambda-{1\over2} \sgn(E_\lambda)],
\end{equation}
and, consequently, its vacuum expectation value takes form
\begin{equation}
\N\equiv<\vac|\hat{\N}|\vac>=-{1\over2}\si
\sgn(E_\lambda)=-{1\over2}\int d^2x\, \tr<\x|\,\sgn(H)|\x>. 
\end{equation}
From general arguments, one could expect that the last quantity
vanishes due to cancellation between the contributions of positive and
negative energy solutions to the Dirac equation (2). Namely this
happens in a lot of cases. That is why every case of a nonvanishing value
of $\N$ deserves a special attention.

Considering the case of the background in the form of singular
magnetic vortex (7) -- (8), one can notice that the contribution
of regular solutions (15) and (16) is cancelled upon summation
over the sign of energy, whereas irregular solution (17) yields a
nonvanishing contribution to $\N$ (20). Defining the vacuum fermion
number density
\begin{equation}
\N_{\x}=-{1\over2}\tr<\x|\,\sgn(H)|\x>, 
\end{equation}
we get
\begin{equation}
\N_\x=-{1\over8\pi}\int\limits_0^\infty dkk\biggl\{
A\biggl({k\over\mu}\biggr)^{2F-1}
\biggl[\Lp+\Lm\biggr]\biggl[J_{-F}^2(kr)+$$
$$+J_{1-F}^2(kr)\biggr]+2\biggl[\Lp-\Lm\biggr]\biggl[J_{-F}(kr)J_F(kr)-
J_{1-F}(kr)J_{-1+F}(kr)\biggr]+$$
$$+A^{-1}\biggl({k\over\mu}\biggr)^{1-2F}\biggl[\Lp+\Lm\biggl]
\biggr[J_F^2(kr)+J_{-1+F}^2(kr)\biggr]\biggr\}, 
\end{equation}
where
\begin{equation}
A=2^{1-2F}{\Gamma(1-F)\over\Gamma(F)}\tan\left(s{\Theta\over2}+
{\pi\over4}\right), 
\end{equation}
\begin{equation}
L_{(\pm)}=2^{-1}\bigl\{\cos(F\pi)\pm\cosh\bigl[(2F-1)\ln({k\over\mu})+
\ln A\bigr]\bigr\}^{-1}. 
\end{equation}
Transforming the integral in Eq.(22), we get the final expression
\begin{equation}
\N_\x=-{\sin(F\pi)\over2\pi^3r^2}\int\limits_0^\infty dw\,
w {K_F^2(w)-K_{1-F}^2(w)\over \cosh[(2F-1)\ln({w\over\mu r})+\ln A]},
\end{equation}
where $K_{\rho}(w)$ is the Macdonald function of order $\rho$.
Vacuum fermion number density (25) vanishes at half integer values
of the vortex flux ($F={1\over2}$) as well as at $\cos\Theta=0$.
Otherwise, at large distances from the vortex we get
\begin{equation}
\N_\x{}_{\stackrel{\ds =}{r\rightarrow
\infty}}-(F-{1\over2}){\sin(F\pi)\over2\pi^2r^2}
\left\{\ba{cc}
(\mu r)^{2F-1}A^{-1}{\ds\Gamma({3\over2}-F)\Gamma({3\over2}-2F)\over\ds
\Gamma(2-F)},& 0<F<{1\over2}\\[0.2cm]
(\mu r)^{1-2F}A{\ds\Gamma(F+{1\over2})\Gamma(2F-{1\over2})\over\ds
\Gamma(1+F)},& {1\over2}<F<1 \\ \ea \right. \,. 
\end{equation}
Integrating Eq.(25) over plane ($x^1,x^2$), we obtain the total
vacuum fermion number
\begin{equation}
\N=-{1\over2}\sgn_0\left[(F-{1\over2})\cos\Theta\right], 
\end{equation}
where
$$\sgn_0(u)=\left\{ \ba{cc}
\sgn(u),& u\neq0\\
0,& u=0\\ \ea \right\}.
$$

\section{Parity Breaking Condensate}

Since twodimensional massless Dirac Hamiltonian (4) anticommutes
with the $\beta$ matrix
\begin{equation}
[H,\beta]_+=0, 
\end{equation}
Dirac equation (2) is invariant under the parity transformation
\begin{equation}
E_\lambda\rightarrow -E_\lambda, \qquad <\x|\lambda>\rightarrow
\beta<\x|\lambda>. 
\end{equation}
However, this invariance is violated by the boundary condition (12), 
unless $\cos\Theta=0$. Consequently, the parity breaking
condensate emerges in the vacuum:
\begin{equation}
\C_\x=<\vac|{1\over2}[\Psi^+(\x,t),\beta\,\Psi(\x,t)]_-|\vac>=
-{1\over2}\tr<\x|\beta\,\sgn(H)|\x>. 
\end{equation}
The contribution of regular solutions (15) and (16) to Eq.(30) is cancelled 
upon summation over the energy sign. Thus, only
the contribution of irregular solution (17) to Eq.(30) survives:
\begin{equation}
\C_\x=-{1\over8\pi}\int\limits_0^\infty {dk\,k}
\biggl\{A\biggl({k\over\mu}\biggr)^{2F-1}\bigl[\Lp+\Lm\bigr]
\bigl[J_{-F}^2(kr) -J_{1-F}^2(kr)\bigr]+$$
$$+2\bigl[\Lp-\Lm\bigr]
\bigl[J_{-F}(kr)J_F(kr)+J_{1-F}(kr)J_{-1+F}(kr)\bigr]+$$
$$
+A^{-1}\biggl({k\over\mu}\biggr)^{1-2F}\bigl[\Lp+\Lm\bigr]
\bigl[J_F^2(kr)-J_{-1+F}^2(kr)\bigr]\biggr\}, 
\end{equation}
where $A$ and $L_{(\pm)}$ are given by Eqs.(23) and (24),
respectively. Transforming the integral in Eq.(31), we get
\begin{equation}
\C_\x=- {\sin(F\pi)\over2\pi^3r^2}\int\limits_0^\infty dw\,
w{K_F^2(w)+K_{1-F}^2(w)\over\cosh[(2F-1)\ln({w\over\mu r})+\ln A]}.
\end{equation}
Evidently, Eq.(32) vanishes if $\cos\Theta=0$. At half integer values
of the vortex flux $(F={1\over2})$, we get
\begin{equation}
\C_\x\big|_{F={1\over2}}=-{\cos\Theta\over 4\pi^2r^2}. 
\end{equation}
At large distances from the vortex we get
\begin{equation}
\C_\x{}_{\stackrel{\ds =}{r\rightarrow
\infty}}-{\sin(F\pi)\over2\pi^2r^2} \left\{\ba{cc} (\mu
r)^{2F-1}A^{-1}{\ds\Gamma({3\over2}-F)\Gamma({3\over2}-2F)\over \ds
\Gamma(1-F)}, & 0<F<{1\over2}\\[0.2cm]
(\mu r)^{1-2F}A {\ds\Gamma(F+{1\over2})\Gamma(2F-{1\over2})\over\ds
\Gamma(F)},& {1\over2}<F<1\\ \ea \right. \,.
\end{equation}
Integrating Eq.(32) over plane ($x^1,x^2$), we obtain the total
vacuum condensate
\begin{equation}
\C\equiv\int d^2x\,\C_\x=-{\sgn_0(\cos\Theta)\over4|F-{1\over2}|}.
\end{equation}
Thus, the total vacuum condensate is infinite at $F={1\over2}$ if
$\cos\Theta\neq0$.

\section{Angular Momentum}

Let $\hat{M}$ be an operator in the first-quantized theory, which
commutes with the Dirac Hamiltonian
\begin{equation}
[\hat{M},H]_-=0. 
\end{equation}
Then, in the second-quantized theory, the vacuum expectation value of the
dynamical variable corresponding to $\hat{M}$ is presented in the form
\begin{equation}
\M=\i d^2x\,\M_\x, 
\end{equation}
where
\begin{equation}
\M_\x=<\vac|{1\over2}\bigl[\Psi^+(\x,t),\hat{M}\,\Psi(\x,t)\bigr]_-|\vac>=
-{1\over2}\tr<\x|\hat{M}\,\sgn(H)|\x>. 
\end{equation}
The commutation relation (36) is the evidence of invariance of the
theory with $\hat{M}$ being the generator of the symmetry transformations. 
Since, in background (7) -- (8), there is invariance with respect to 
rotations around the location of the vortex, one can take $\hat{M}$ as the
generator of rotations -- the operator of angular momentum in the
first-quantized theory (see \cite{SitR96} for more details):
\begin{equation}
\hat{M}=-i\x\times\bpar+{1\over2}s\beta. 
\end{equation}
Note that the eigenvalues of operator $\hat{M}$ (39) on spinor
functions are half integer.

Decomposing Eq.(39) into the orbital and spin parts, we get in the
second-quantized theory
\begin{equation}
\M_\x=\L_\x+\S_\x, 
\end{equation}
where
\begin{equation}
\L_\x={1\over2}\tr<\x|(i\x\times\bpar)\,\sgn(H)|\x> 
\end{equation}
and
\begin{equation}
\S_\x=-{1\over4}s\,\tr<\x|\beta\,\sgn(H)|\x>. 
\end{equation}
Since vacuum spin density (42) is related to vacuum condensate (30),
\begin{equation}
\S_\x={1\over2}s\,C_\x, 
\end{equation}
there remains only vacuum orbital angular momentum density (41) to
be considered.

The contribution of regular solutions (15) and (16) 
is cancelled upon summation over the energy sign, whereas the
contribution of irregular solution (17) yields:
\begin{equation}
\L_\x=-{1\over8\pi}\int\limits_0^\infty\, {dk\,k} 
\biggl\{ A\biggl({k\over\mu}\biggr)^{2F-1}\bigl[
\Lp+\Lm\bigr]\bigl[n_0 J_{-F}^2 (kr)
+(n_0+s)J_{1-F}^2(kr)\bigr]+$$
$$+2\bigl[\Lp-\Lm\bigr]\bigl[n_0J_{-F}(kr)J_F(kr)
-(n_0+s)J_{1-F}(kr)J_{-1+F}(kr)\bigr]+$$
$$+A^{-1}\biggl({k\over\mu}
\biggr)^{1-2F} \bigl[\Lp+\Lm\bigr]\bigl[n_0J_F^2(kr)+
(n_0+s)J_{-1+F}^2(kr)\bigr]\biggr\}.
\end{equation}
Transforming the integral in Eq.(44), we get
\begin{equation}
\L_\x=-{\sin(F\pi)\over2\pi^3r^2}\int\limits_0^\infty dw\, w
{n_0K_F^2(w)-(n_0+s)K_{1-F}^2(w)\over \cosh[(2F-1)\ln({w\over\mu
r})+\ln A]}. 
\end{equation}
Summing Eqs.(45) and (46), taking into account Eqs.(32) and
(11), we obtain the following expression  for the vacuum angular
momentum density in background (7) -- (8):
\begin{equation}
\M_\x=\biggl(\io\F\ic+{1\over2}\biggr)\,\N_\x, 
\end{equation}
where vacuum fermion number density $\N_\x$ is given by Eq.(25).
Consequently, the total vacuum angular momentum takes form (see Eq.(27))
\begin{equation}
\M=-{1\over2}\biggl(\io\F\ic+{1\over2}\biggr)\,
\sgn_0\bigl[(F-{1\over2})\cos\Theta\bigr]. 
\end{equation}
Thus, if the vacuum angular momentum could become somehow 
detectable, then this would provide us with a unique
explicit evidence in favour of physical effects that depend
essentially both on integer and fractional parts of the flux 
of a singular magnetic vortex.

\noindent{\bf Acknowledgements.}
The work was supported by the U.S. Civilian Research and Development Foundation 
(CRDF grant UP1-2115).

\end{document}